\documentclass[12pt,a4paper]{article}

\usepackage{authblk}
\usepackage{graphicx}
\usepackage{multirow}
\usepackage{wasysym}
\usepackage{rotating}

\begin{document}
\title{The Quest for $\mu \to e \gamma$ and its Experimental Limiting Factors at Future High Intensity Muon Beams}


\author[1]{Francesco Renga}
\author[1,2]{Gianluca Cavoto}
\author[3,4]{Angela Papa}
\author[5]{Emanuele Ripiccini}
\author[1]{Cecilia Voena}

\affil[1]{Istituto Nazionale di Fisica Nucleare, Sez. di Roma, P.le A. Moro 2, 00185 Roma, Italy}
\affil[2]{``Sapienza'' Universit\`a di Roma, Dipartimento di Fisica, P.le A. Moro 2, 00185 Roma, Italy}
\affil[3]{Dipartimento di Fisica, Università di Pisa, and INFN sez. Pisa, Largo B. Pontecorvo 3, 56127 Pisa, Italy}
\affil[4]{Paul Scherrer Institut, 5232 Villigen, Switzerland}
\affil[5]{Universit\'e de Gen\`eve, D\'epartement de physique nucl\'eaire et corpusculaire, 24 Quai Ernest-Ansermet, 1211 Gen\`eve, Switzerland}

\maketitle







\abstract{The search for the Lepton Flavor Violating decay $\mu \to e \gamma$ exploits the most intense continuous muon beams, which can currently deliver
$\sim 10^8$ muons per second. In the next decade, accelerator upgrades are expected in various facilities, making it feasible to have continuous beams with 
an intensity of $10^9$ or even $10^{10}$ muons per second. We investigate the experimental limiting factors that will define the ultimate performances, and 
hence the sensitivity, in the search for $\mu \to e \gamma$ with a continuous beam at these extremely high rates. We then consider some conceptual detector 
designs and evaluate the corresponding sensitivity as a function of the beam intensity.}









\newcommand{\meg}{\ensuremath{\mu^+ \to e^+ \gamma}}



\section{Introduction}

In the Standard Model (SM), if the neutrino masses are neglected, three families (or \emph{flavors}) of leptons exist, and in any process the number of leptons of 
each family is separately conserved. Lepton flavor conservation is anyway an accidental symmetry: a mere consequence of the particle content of the model, following
namely from the absence of right handed neutrinos. Actually, this symmetry is not exact, as already demonstrated by the discovery of neutrino oscillations,but the
impact on charged lepton decays is negligible, giving for instance a predicted branching ratio (BR) for the $\mu \to e \gamma$ decay around $10^{-54}$,
well below the current experimental limit, $BR(\mu \to e \gamma) < 4.2 \times 10^{-13}$~\cite{meg_analysis}. 

On the other hand, the accidental nature of this symmetry makes it very sensitive to new physics (NP) processes beyond the SM. In many models lepton flavor violation (LFV)
in the charged lepton sector arises in a measure that is already strongly constrained by the present experimental limits.

These features make the search for LFV very attractive, because negative results are able to strongly constrain the development of NP models, while an observation
would be an unambiguous evidence of physics beyond the SM, with no theoretical uncertainty.

In this work we investigate the potential of the next generation of searches for the LFV decay $\mu \to e \gamma$, in the view of the possible availability of high intensity
muon beams, delivering a number of muons per second up to two orders of magnitude larger than what is presently possible at the Paul Scherrer Institut (PSI, Switzerland),
where the most intense continuous muon beam line in the world is operated, with up to $10^8$ muons per second. Projects to reach a muon beam rate 
of $10^9$ or even $10^10$ muons per second are under considerations at PSI~\cite{HiMB} and elsewhere~\cite{MuSIC,PIP-II}. In this kind of facilities, muons come
from the decay of pions produced by a proton beam impinging on a fixed target. At PSI an high intensity muon beam line (HiMB) is studied, that should be able 
to increase by a factor $> 4$ the muon capture efficiency at the production target, thanks to a new design of the solenoid magnets used to convey the muons into
the beam line, and by a factor $\sim 7$ the transport efficiency from the production target to the experimental halls, thanks to an improved beam optics. The main
limitation at PSI comes from the fact that the proton beam need to be mostly preserved to serve a neutron spallation source downstream of the muon production
target. Hence, a thin target is used and only 18\% of the original, 2~mA beam is used to produce muons. At RCNP (Japan), the MuSIC project aims to use
a thicker target in order to get a similar production rate with a much lower proton beam intensity. The target will be surrounded by an intense solenoidal 
magnetic field in order to capture pions and muons with a large solid angle acceptance, and a magnetic field adiabatically changing from 3.5 T at the center of the 
target to 2 T at the exit of the capture solenoid will reduce the angular divergence of the beam and hence increase the acceptance of the solenoidal muon 
transport beam line. The tests already performed showed that $\sim 10^6$ muons per Watt of proton beam power can be produced. Some studies are also on going
for the production of continuous muon beams in the context of the PIP-II project at Fermilab (USA).

\section{Materials and Methods}

Our discussion of the sensitivity reach of future experiments looking for $\mu \to e \gamma$ considers the typical features of this kind of searches. 
First of all, positive muons are used in these experiments, in order to avoid their capture in the target nuclei, which would distort the energy spectrum 
of the decay products. In order to get an intense muon beam with low contamination of pions and electrons, the beam lines are tuned to transport particles 
of about 28~Mev/c of momentum, corresponding to muons produced by pions decaying at rest just at the surface of the production target. Muons of such a 
low momentum can be stopped on thin targets, and typically a few hundred microns of plastic material are sufficient. 
The muons decay at rest and the two-body $\mu \to e \gamma$ kinematics is exploited, by looking for a positron and a photon emitted back to back, 
with equal energies (neglecting the electron mass), $E_e = E_\gamma = m_\mu c^2/2 \sim 52.8$~MeV.

There is a physics background coming from the radiative muon decay (RMD),  $\mu^+ \rightarrow e \gamma \nu_e  \bar{\nu_{\mu}}$, when 
the two neutrinos carry off little energy. The other  is due to the  accidental coincidence of a positron from a Michel muon decay,  
$\mu^+ \rightarrow e^+ \nu_e  \bar{\nu_{\mu}}$, with a high energy photon, whose source might be either a RMD, the 
annihilation-in-flight  (AIF) of a positron in a Michel  decay or the bremsstrahlung from a positron. The rate of accidental coincidences, goes with the
square of the muon beam rate $\Gamma_\mu$, and hence it becomes dominant over RMDs at very high muon beam rates. In order to discriminate against
accidental coincidences, the difference between the positron and photon emission time, $T_{e\gamma}$, is also required to be zero.

If a signal region is defined in the parameter space given by the photon and positron energies, the relative stereo angle $\Theta_{e\gamma}$ and the
relative time, with dimensions proportional to the detector resolutions $\delta E_e$, $\delta E_\gamma$$\delta \Theta_{e\gamma}$ and $T_{e\gamma}$, 
the accidental rate is found to be~\cite{kuno-okada}:
\begin{equation}
\label{eq:acc_rate}
\Gamma_{\mathrm{acc}} \propto \Gamma_\mu^2 \cdot \delta E_e \cdot (\delta E_\gamma)^2 \cdot \delta T_{e\gamma} \cdot (\delta \Theta_{e\gamma})^2
\end{equation}
It indicates that, if a significative background yield is expected during the lifetime of the experiment, a further increase of the beam intensity is useless, because
the sensitivity, depending on the ratio of the signal yield over the square root of the background yield remains constant, unless the resolutions are improved
in such a way that the background yield becomes negligible. Equivalently, given an experimental setup running for a given time, there is an ideal muon beam rate:
the one giving only very few expected background events. 

We analyzed in detail the experimental factor that will limit the sensitivity of the future searches for $\mu \to e \gamma$. We considered different detector options,
both for the positron and the photon, and performed simulations to determine the performances that could be reasonably reached. We made this exercise considering
different experimental approaches. 

In particular, we studied the two different techniques that can be used to detect the photon. In a calorimetric approach, a fast and luminous inorganic scintillating material 
is used to measure the photon energy, time and position at the detector. Alternatively, thin layers of dense material
can be used to produce a photon conversion to $e^+e^-$, and the two charged particles are tracked in a magnetic field to determine the photon energy and
conversion point. The first approach, used in the last decades by the CrystalBox~\cite{crystal-box} and MEG~\cite{meg-detector} experiments,
gives a relatively high efficiency ($> 60\%$ in MEG), while the resolutions are determined by the physical properties of the scintillating material. 
The MEG experiment, and its upgrade MEG-II (currently under construction) have a LXe calorimeter, with an homogeneous volume of $900~\ell$, giving very good 
energy ($\sim 2\%$) and time ($\sim 60$~ps) resolutions. The second approach, adopted for the MEGA experiment~\cite{mega}, suffers from a very low efficiency 
(a few percent per conversion layer) due to the low conversion probability, but allows to reach extremely good resolutions, which according to the discussion 
above can allow to exploit a higher beam rate to recover the loss of efficiency. Neither technique provides a precise determination of the photon direction. It is more 
precisely determined by tracking the positron back to the target, assuming that the photon comes from the same place, and taking the line joining this point to the photon 
conversion point as the photon direction. Nonetheless, the conversion technique gives some information about the photon direction, as the combination of the directions of 
the $e^+e^-$ pair. This supplementary information can be used to require the photon and the positron to come approximately from the same point, and it 
helps to reduce the accidental background.

Second, we considered the best performances that could be reached in the positron reconstruction, which is typically carried on with a magnetic spectrometer,
which provides high efficiency and the best resolutions in momentum and direction.

We also considered the impact of the target and other detector materials. Due to the low positron momentum, the multiple Coulomb scattering (MS)
plays a dominant role in the determination of the positron kinematics, and the target itself, as thin as it can be, still gives non negligible contributions. Moreover,
materials on the positron trajectory increase the probability of producing AIF photons and hence the accidental background.

\begin{figure}[h]
\centering
\resizebox{0.75\textwidth}{!}{
 \includegraphics{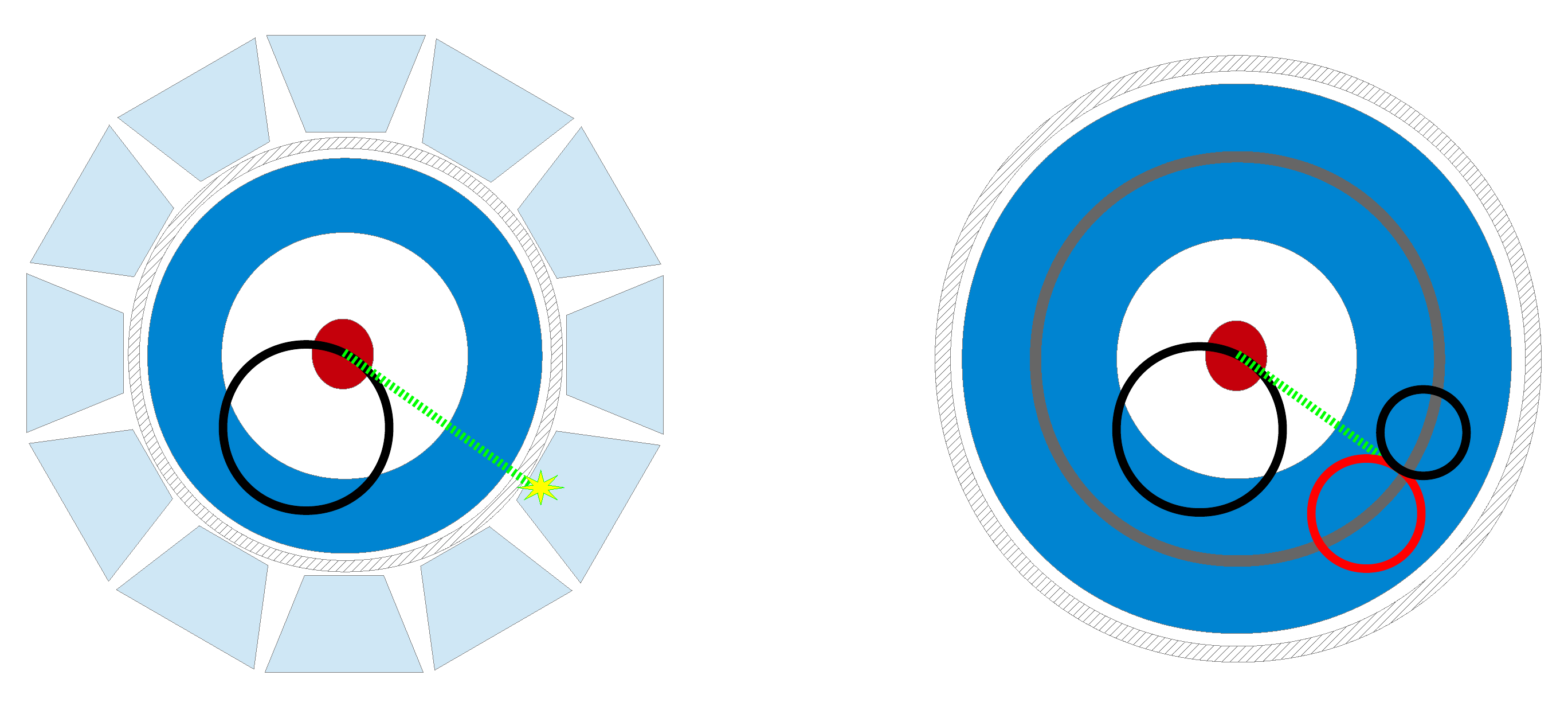}
}
\caption{Conceptual detector designs exploiting the calorimetric (left) or conversion (right) technique for the photon detection, and a tracking approach in a magnetic 
field for the positron reconstruction. Muons are stopped in a target (dark red ellipse) at the center of the magnet. Positron tracks from the muon decays (in black) 
are reconstructed in a tracking detector (dark blue), photons (in green) either produce a shower in a calorimeter (light blue) or are converted by a thin layer of 
high-Z material (in gray) into an electron-positron pair (in red and black, respectively) which is then reconstructed by an outer tracking detector. The magnet coil (hatched
area) surrounds the tracking detectors. }
\label{fig:sketch_basics}
\end{figure}

In Figure~\ref{fig:sketch_basics} the conceptual design of a detector searching for $\mu \to e \gamma$ is shown, for the calorimetric and the conversion techniques.

\section{Results}

\subsection{Experimental limiting factors}

\subsubsection{Efficiency}
\label{sec:eff}

The signal efficiency is determined by the positron and photon reconstruction efficiencies, $\epsilon_e$ and  $\epsilon_{\gamma}$. 
First of all, the geometrical acceptance has to be considered. It is typically constrained by the cost of the photon detector. 
The MEG experiment, for instance, only had a 10\% acceptance, limited by the angular coverage of the (very expensive) LXe calorimeter. 
Though mitigated, this point could be relevant also for the innovative crystals we will discuss in Sec.~\ref{sec:Eg}.

If the calorimetric technique is used, the efficiency is limited by the number of photons converting before reaching the calorimeter,
typically in the material of the magnet of the positron spectrometer. A reconstruction efficiency of $\sim 60\%$ was obtained in MEG.

\begin{figure}[h]
\centering
\resizebox{0.5\textwidth}{!}{
  \includegraphics{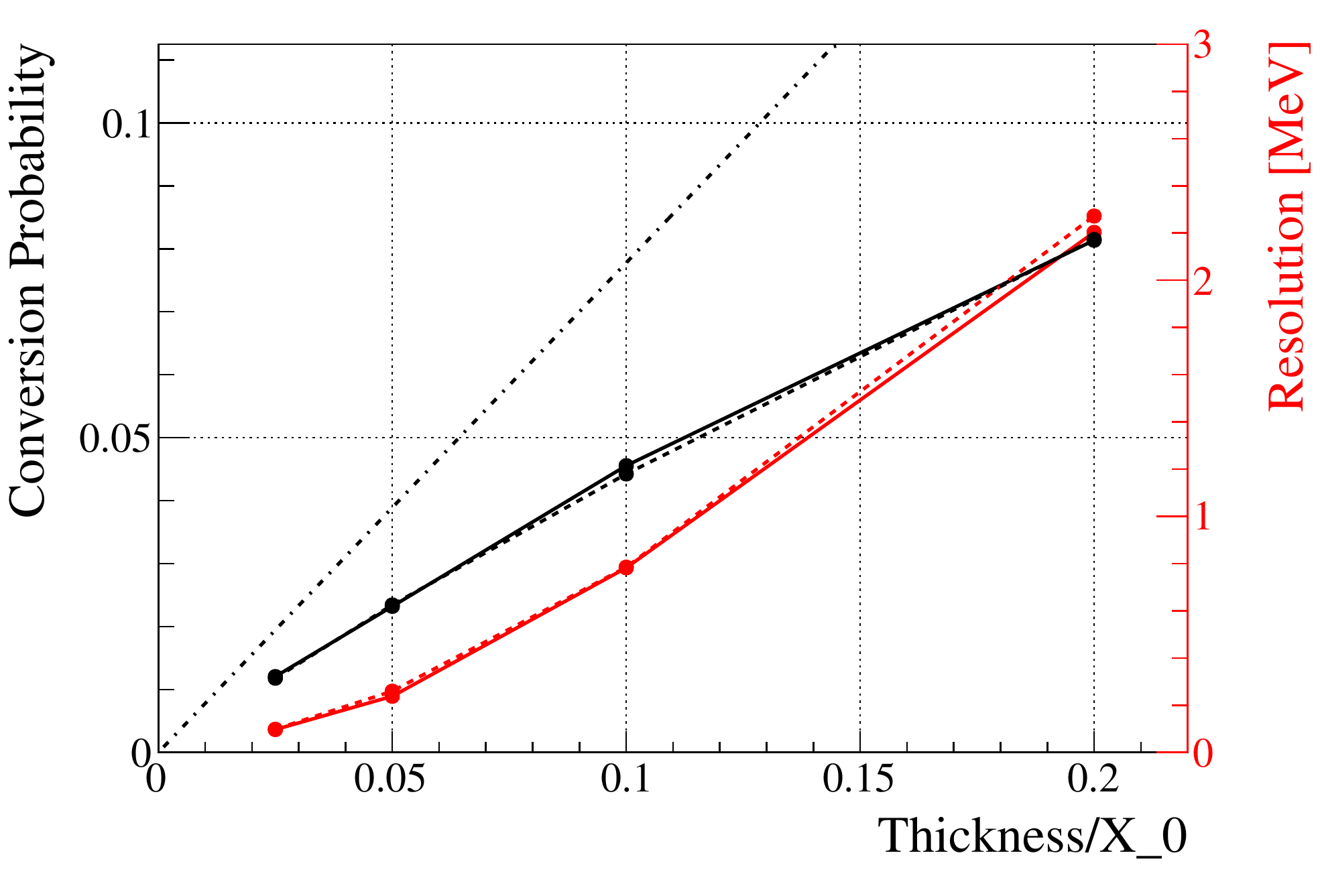}
}
\caption{The conversion efficiency (black, left axis) and the contribution to the energy resolution from the energy loss in the converter (red, right axis), for Lead (full lines) 
and Tungsten (dashed lines), as a function of the converter thickness (in units of radiation length). The dash-dotted line shows the asymptotic conversion probability, 
$7/9$ times the thickness in units of radiation length.}
\label{fig:Egamma_reso_simple}
\end{figure}

If the photon conversion technique is adopted, thin converters are needed in order to preserve very good resolutions. It implies in turn 
a few percent  $\epsilon_{\gamma}$. In Fig.~\ref{fig:Egamma_reso_simple} the conversion probability for 52.8~MeV photons in lead and 
tungsten for different thicknesses are shown.

Concerning the positron from the muon decay, tracking in a spectrometer usually provides very large $\epsilon_{e}$.

\subsubsection{Photon energy}
\label{sec:Eg}

In a calorimetric approach, the $E_\gamma$  resolution is dominated by the photon statistics. Hence, the light yield determines the choice of the scintillator to be 
used, along with the fast response that is needed in order to reach a very good time resolution. Table~\ref{tab:scintillators} summarizes the 
relevant properties of some  state-of-the-art scintillating materials. LaBr$_3$(Ce) crystals are a good candidate for future experiments, thanks to
the high light yield, which should guarantee a good energy resolution and the low decay time, which is necessary to get a very good time resolution.

\begin{table}[h]
\caption{\label{tab:scintillators} Properties of state-of-the-art scintillators relevant for the application on \meg~searches.}
\centering
\begin{tabular}{lccc}
\hline
\textbf{Scintillator} & \textbf{Density} & \textbf{Light Yield} & \textbf{Decay Time}\\
& \textbf{[g/cm$^3$]} & \textbf{[ph/keV]} & \textbf{[ns]}\\
\hline
LaBr$_3$(Ce) & 5.08 & 63 & 16 \\
LYSO & 7.1 & 27 & 41 \\
YAP & 5.35 & 22 & 26 \\
LXe & 2.89 & 40 & 45 \\
NaI(Tl) & 3.67 & 38 & 250 \\
BGO & 7.13 & 9 & 300 \\
\hline
\end{tabular}
\end{table}

I pair conversion is used, the limiting factor of the $E_\gamma$ resolution is the interaction of the $e^+e^-$ pair within the material of the photon converter itself.
The energy loss fluctuation predominantly contributes to the resolution, since $E_\gamma$ is estimated as the sum of the $e^+$ and $e^-$ energies. 
We performed simulations with GEANT4~\cite{geant4} showing that a 280~$\mu$m Pb layer ($\sim 5\%~X_0$) would give a resolution of $\sim 240$~keV 
in the limit of perfect tracking of the $e^+e^-$ pair. Figure~\ref{fig:Egamma_reso_simple} also show the contribution of the material effects to the resolution.


\subsubsection{Positron energy}
\label{sec:Ee}

The positron energy and positron angular resolutions in a spectrometer are ultimately determined by MS and energy loss fluctuations. For this reason, gaseous 
detectors give the best performances and have been used in the latest experiments. A silicon vertex tracker is used for the search of $\mu^+ \to e^+ e^+ e^-$ 
by the Mu3e  Collaboration~\cite{mu3e-tracker}, and a similar design has been suggested for future \meg~searches \cite{caltech}, considering
that very thin sensors are now available, with a thickness of 50~$\mu$m Si + 25~$\mu$m Kapton per layer~\cite{hvmaps}, corresponding to $\sim 10^{-3}$ radiation
lengths per layer. On the other hand, the complete drift-chamber spectrometers of MEG or MEG-II amount to less than $3 \times 10^{-3}$ radiation lengths 
over  the whole track length within the tracking volume, nonetheless material effects gave a significant contribution in MEG and will almost be  dominant in MEG-II. 
It clearly indicates that more than a few silicon layers cannot be used: indeed, simulations~\cite{caltech} point toward $E_e$ resolutions of $\sim$~200~keV, which are 
not competitive with what can be obtained with gaseous detectors~\cite{meg2}.

\subsubsection{Relative angle $\Theta_{e\gamma}$ }
\label{sec:Thetaeg}

The relative angle $\Theta_{e\gamma}$ is measured  by combining the positron angle, the photon conversion point and the positron vertex on the target,
which has to be as thin as possible to reduce the MS affecting the positron angle measurement. On the other hand it has to be thick enough to provide a good 
stopping power for muons. A good compromise has been obtained by slanting the target with respect to the beam axis (in MEG the target normal vector makes 
an angle $\alpha \sim 70^\circ$ with the beam axis, which will be increased to $76^\circ$ in MEG-II). In this configuration, the effective thickness seen by muons 
is magnified by a factor $1/\cos(\alpha) \sim 3$, while positrons emitted at the center of the detector acceptance ($90^\circ$ with respect to the beam axis) see a 
thickness magnified only by a factor $1/\sin(\alpha) \sim 1.06$. Nonetheless, if a 90\% stopping efficiency is required, simulations suggest that the contribution 
to the angular resolutions is always larger than 3~mrad, even with the material (Beryllium) giving the best performances. Also, reducing the thickness by accepting
a loos off stopping efficiency is not feasible: the target has to be placed at the Bragg peak to have a reasonable stopping efficiency, and survived muons would
decay in the gas just after the target, giving a contribution to the background without increasing the signal rate. Hence, a positron angle resolution better than 
a few mrad cannot be obtained with conventional techniques.

With the photon conversion technique, the photon conversion point can be measured very precisely, essentially with the single hit resolution of the 
$e^+e^-$ tracker. As a consequence, the photon angle resolution is completely dominated by the positron vertex resolution. With calorimetry, 
the granularity of the detector determines the resolution, but it is reasonable to assume a resolution below 1~cm. In both cases, the positron
angle resolution is dominant and defines the ultimate $\Theta_{e\gamma}$ resolution.

\subsubsection{Relative time $T_{e\gamma}$ }
\label{sec:Teg}

A good $T_{e\gamma}$ resolution has been guaranteed in the MEG experiments by the use of scintillation detectors placed at the end of the 
positron trajectory, in combination with the good time resolution of the LXe calorimeter. Replicating these performances in future experiments
will require either the use of a calorimetric approach with very fast crystals or the inclusion of scintillators on the $e^+e^-$ trajectory if the
photon conversion is used.

\subsubsection{Summary}

Tab.~\ref{tab:limits} shows a summary of the limiting factors for the efficiency and resolutions of future \meg~searches, as derived
from simulations and analysis of past experiments. More details can be found in~\cite{meg-future}.

\begin{sidewaystable}
\caption{\label{tab:limits} Limiting factors for the efficiency and resolutions of future \meg~searches.}
\centering
\begin{tabular}{lccc}
 & \multicolumn{2}{c}{\bf Typical figure} & \multirow{2}{*}{\bf Comments} \\
 & \emph{Calorimetry} & \emph{$\gamma$ Conversion} & \\
\cline{2-3}
\multicolumn{4}{l}{\emph{Efficiency}}\\
\hline  
\hline  
Material budget & 0.5 $\sim$ 0.9 & -- & magnet coil \\
Pair production & -- & 0.02 $\sim$ 0.04 & 0.05 $\sim$ 0.1 $X_0$\\
Minimum $e^+e^-$ energies &  -- & 0.8  & $E_{e^+}, E_{e^-} > 5$~MeV \\
 & & & \\
\multicolumn{4}{l}{\emph{Photon Energy Resolution}}\\
\hline 
\hline 
Energy loss & -- & 250 $\sim$ 800~keV & 0.05 $\sim$ 0.1 $X_0$\\ 
Photon Statistics \& segmentation & 800~keV & -- & \\
 & & & \\
\multicolumn{4}{l}{\emph{Positron Energy Resolution}}\\
\hline 
\hline 
Energy loss & \multicolumn{2}{c}{15~keV} & \\  
Tracking \& MS & \multicolumn{2}{c}{100~keV} & \\ 
 & & & \\
\multicolumn{4}{l}{\emph{Relative Angle Resolution}} \\
\hline 
\hline
MS on target & \multicolumn{2}{c}{2.6~/~2.8~mrad ($\theta_{e\gamma}$ / $\phi_{e\gamma}$)} & \\ 
MS on gas \& walls & \multicolumn{2}{c}{3.3~/~3.3~mrad ($\theta_{e\gamma}$ / $\phi_{e\gamma}$)} & \multirow{2}{*}{$R_e = 20$~cm, $R_\gamma = 30~$cm, B = 1~T} \\
Tracking & \multicolumn{2}{c}{6.0~/~4.5~mrad ($\theta_{e\gamma}$ / $\phi_{e\gamma}$)} & \\
Alignment & \multicolumn{2}{c}{$< 1$ mrad} & $<$100~$\mu$m target alignment\\
 & & & \\
\hline
\end{tabular}
\end{sidewaystable}

\subsection{Sensitivity reach}
\label{sec:sens}

We considered a conceptual \meg~detector based on the photon conversion technique. In this design, a target identical to the one of 
MEG-II is surrounded by a cylindrical gaseous positron tracker. Externally, a Lead conversion layer is placed, with a $0.1~X_0$ thickness. 
Behind it, another gaseous detector is used as an $e^+e^-$ pair spectrometer.

Optionally, a small gaseous or two-layer solid state detector can be considered as a vertex tracker to improve the determination of 
the positron angles and the muon decay point.

Everything is immersed in a magnetic field. The signal positron curls before reaching the converter layer and 
finally reaches a set of scintillators for positron timing, while at least one of the tracks from the photon conversion goes 
through the whole $e^+e^-$pair spectrometer and reaches another set of scintillators for the photon timing.
 
We estimated the expected performances of such a detector, assuming that the ultimate resolutions of Table~\ref{tab:limits} can be reached with
an incremental improvement of the present experimental techniques. We consider two scenarios for the inner vertex detector.
In the first, conservative one, the only improvement comes from having the first measured point which is closer to the target, 
while the momentum and angular resolutions are still dominated by the extended tracker, and the angular resolution is deteriorated 
by the MS in the inner wall of the TPC or the inner layer of the silicon vertex tracker. In the second, optimistic one, the vertex detector makes also
the tracking contribution to the angular resolution negligible. This resolution is then completely determined by material effects before and inside the first 
layer of the inner vertex detector. A summary of the expected performances can be found in Tab.~\ref{tab:perf_ee_1}  and ~\ref{tab:perf_ee_2}. 
It is evident that a silicon vertex detector cannot help, because the MS in the first layer of such a detector negates the advantage of having a very good 
determination of the track angle between the first and the subsequent layers.

\begin{table}[h!t]
\begin{center}
\caption{\label{tab:perf_ee_1}  Expected performances (efficiency and resolutions)  for a basic design with different options as discussed in the text.}
\begin{tabular}{lcc}
{\bf Observable} &  \emph{one photon } &  \emph{ photon }  \\
                          &  \emph{conversion layer} &  \emph{  calorimeter}  \\
\hline
$T_{e\gamma}$ [ps] & 60  & 50 \\
$E_e$ [keV] & 100   & 100 \\
$E_\gamma$ [keV] & 320   & 850 \\
Efficiency [\%] & 1.2   & 42 \\
\hline
\end{tabular}
\end{center}
\end{table}

\begin{table}[h!t]
\begin{center}
\caption{\label{tab:perf_ee_2}  Angular resolutions for different types of a vertex detector. A conservative estimate is given in parenthesis.}
\begin{tabular}{lcc}
                          &  $\theta_{e\gamma}$  [mrad] &  $\phi_{e\gamma}$  [mrad] \\
\hline
None   & 7.3  & 6.2   \\
  TPC  &  3.5 (6.1)  & 3.8 (4.8)   \\
   Silicon  &  8.0 (6.3) & 7.4 (6.9) \\
\hline
\end{tabular}
\end{center}
\end{table}

A conceptual \meg~experiment based on calorimetry could have a design very similar to the one above for the central part of the detector, 
but the external  $e^+ e^-$ pair tracker would be replaced by a scintillation detector placed outside of the magnet. With LaBr$_3$(Ce) crystals, the calorimeter could be 
about 20~cm deep and the performance summarized in Tab.~\ref{tab:perf_ee_1}  and ~\ref{tab:perf_ee_2}  could be reached. Here we assume 
that the photon conversion point can be still determined with a negligible resolution compared to the positron vertex resolution.

With these performances and 100 weeks of data taking (3 to 4 years at PSI), with muon rates from $10^{8}$ to $10^{10}$ muons per second, and
assuming the same photon background rate of the MEG experiment (scaled linearly with the muon beam intensity), we could estimate the
the expected sensitivity of the experiment according to a frequentistic approach~\cite{feldman-cousins}.

\begin{figure*}
\centering
\resizebox{0.65\textwidth}{!}{
  \includegraphics{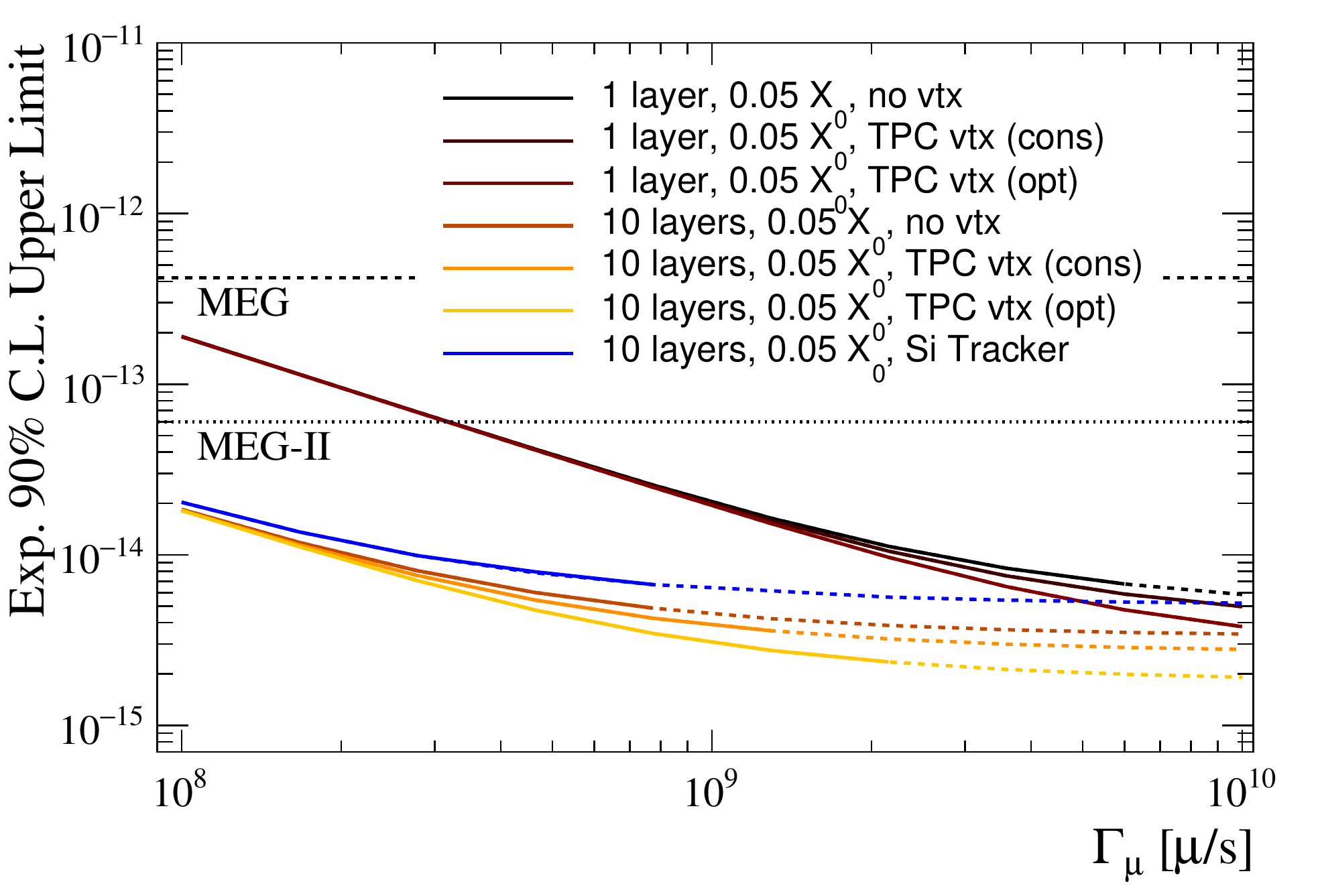}
}
\caption{Expected 90\% C.L. upper limit on the Branching Ratio of \meg~in different scenarios for a 3-year run. A few different designs based on the 
photon conversion technique are compared, including the TPC vertex detector option in the conservative and optimistic hypotheses. The lines turn from continuous to 
dashed when the number of background events exceeds 10. The horizontal dashed and dotted lines show the current MEG limit and the expected MEG-II sensitivity, 
respectively.}
\label{fig:sens1}
\end{figure*}

\begin{figure*}
\centering
\resizebox{0.65\textwidth}{!}{
  \includegraphics{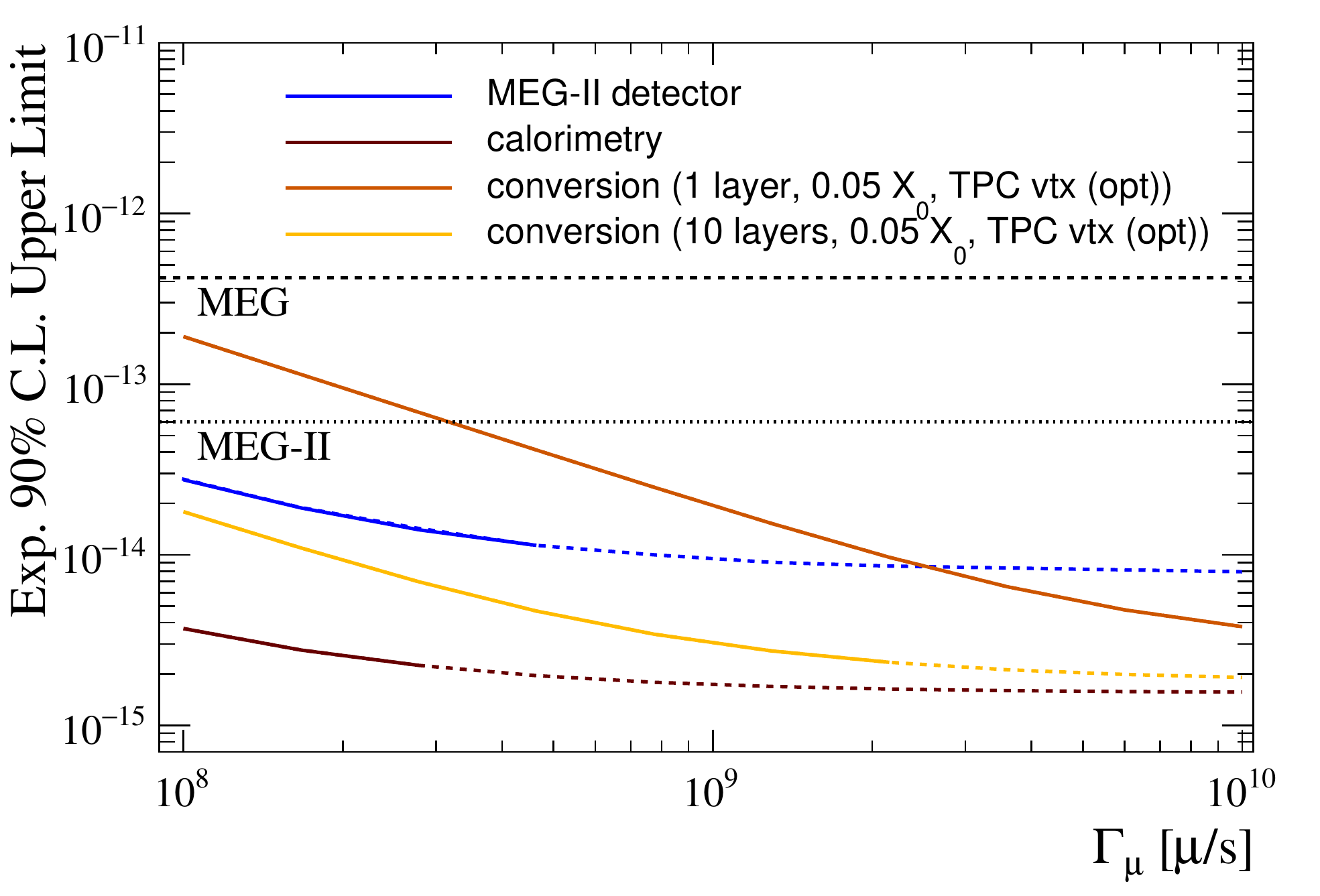}
}
\caption{Expected 90\% C.L. upper limit on the Branching Ratio of \meg~in different scenarios for a 3-year run. Calorimetry and the photon conversion technique are 
compared. The lines turn from continuous to dashed when the number of background events exceeds 10. The horizontal dashed and dotted lines show the current MEG 
limit and the expected MEG-II sensitivity.}
\label{fig:sens2}
\end{figure*}

Figures~\ref{fig:sens1} and \ref{fig:sens2} show the expected sensitivity to the \meg~decay as a function of the beam intensity in different scenarios. We also considered
the possibility of having multiple conversion layers. In this case, the preservation of a good time resolution requires
the inclusion of thin and fast detectors in the conversion layer itself~\cite{meg-future}.

\section{Discussion}

The search for LFV is one of the most promising field in the quest for NP. The present limit on $\mu \to e \gamma$ by the MEG 
collaboration already strongly constrains the NP models and an improvement of one order of magnitude is expected with MEG-II. 
We investigated some long term prospects for the $\mu \to e \gamma$ search. Our estimates show that a 3-year run with an 
accelerator delivering 
around $10^9$ muons per second could allow to reach a sensitivity of a few $10^{-15}$ (expected 90\% upper limit 
on the $\mu \to e \gamma$~BR), with poor perspectives of going below $10^{-15}$ even with $10^{10}$ muons per second. 
Below $5 \times 10^8$ muons per second, the calorimetric approach needs to be used in order to reach this target. 
If a muon beam rate exceeding $10^9$ muons per second is available, the much cheaper photon conversion option 
would be recommended and would provide similar sensitivities.

The sensitivity would be eventually limited by the fluctuations of the interaction of the particles with the detector materials: 
this indicates that a further step forward in the search for $\mu \to e \gamma$~would require a radical  rethinking of the 
experimental concept.

\vspace{6pt} 


\begin{thebibliography}{999}
\bibitem{meg_analysis}
 A.~M.~Baldini {\it et al.} [MEG Collaboration],
  Search for the lepton flavour violating decay $\mu ^+ \rightarrow \mathrm {e}^+ \gamma $ with the full dataset of the MEG experiment,
  Eur.\ Phys.\ J.\ C {\bf 76} (2016) no.8,  434.

\bibitem{HiMB}
P.~R.~Kettle, contribution to Future Muon Sources 2015, University of Huddersfield, United Kinkdom;
A.~Knecht, contribution to SWHEPPS2016, Unter{\"a}geri, Switzerland;

\bibitem{MuSIC}
  S.~Cook {\it et al.},
  Delivering the world's most intense muon beam,
  Phys.\ Rev.\ Accel.\ Beams {\bf 20} (2017) no.3,  030101.

\bibitem{PIP-II}
  V.~Lebedev, ed. [PIP-II Collaboration],
  The PIP-II Reference Design Report,
  FERMILAB-DESIGN-2015-01.

\bibitem{kuno-okada}
  Y.~Kuno and Y.~Okada,
  Muon decay and physics beyond the standard model,
  Rev.\ Mod.\ Phys.\  {\bf 73} (2001) 151.
  
\bibitem{crystal-box}
R.~D.~Bolton {\it et al.} [Crystal Box Collaboration],
  Search for Rare Muon Decays with the Crystal Box Detector,
  Phys.\ Rev.\ D {\bf 38} (1988) 2077.

\bibitem{meg-detector}
J.~Adam {\it et al.} [MEG Collaboration],
  The MEG detector for ${\mu}^+\to e^+{\gamma}$ decay search,
  Eur.\ Phys.\ J.\ C {\bf 73} (2013) no.4,  2365.

\bibitem{mega}
M.~L.~Brooks {\it et al.} [MEGA Collaboration],
  New limit for the family number nonconserving decay mu+ ---> e+ gamma,
  Phys.\ Rev.\ Lett.\  {\bf 83} (1999) 1521.

\bibitem{geant4}
  S.~Agostinelli {\it et al.} [GEANT4 Collaboration],
  GEANT4: A Simulation toolkit,
  Nucl.\ Instrum.\ Meth.\ A {\bf 506} (2003) 250.

\bibitem{mu3e-tracker}
  N.~Berger {\it et al.},
  A Tracker for the Mu3e Experiment based on High-Voltage Monolithic Active Pixel Sensors,
  Nucl.\ Instrum.\ Meth.\ A {\bf 732} (2013) 61

\bibitem{caltech}
  C.~h.~Cheng, B.~Echenard and D.~G.~Hitlin,
  The next generation of $\mu\ -> e \gamma$ and $\mu\ -> 3e$ CLFV search experiments,
  arXiv:1309.7679 [physics.ins-det].

\bibitem{hvmaps}
 I.~Peric {\it et al.},
  High-voltage pixel detectors in commercial CMOS technologies for ATLAS, CLIC and Mu3e experiments,
  Nucl.\ Instrum.\ Meth.\ A {\bf 731} (2013) 131.

\bibitem{meg2}
  A.~M.~Baldini {\it et al.} [MEG II Collaboration],
  The design of the MEG II experiment,
  Eur.\ Phys.\ J.\ C {\bf 78}, no. 5, 380 (2018)

\bibitem{meg-future}
  G.~Cavoto, A.~Papa, F.~Renga, E.~Ripiccini and C.~Voena,
  The quest for $\mu \rightarrow e \gamma $ and its experimental limiting factors at future high intensity muon beams,
  Eur.\ Phys.\ J.\ C {\bf 78}, no. 1, 37 (2018).

\bibitem{feldman-cousins}
  G.~J.~Feldman and R.~D.~Cousins,
  A Unified approach to the classical statistical analysis of small signals,
  Phys.\ Rev.\ D {\bf 57} 3873 (1998).


\end{thebibliography}
\end{document}